\documentclass[twocolumn,showpacs,preprintnumbers,amsmath,amssymb,prl,letterpaper,floatfix,nofootinbib,superscriptaddress]{revtex4-1}
\usepackage{graphics}
\usepackage{graphicx}
\usepackage{epsfig}
\usepackage[usenames]{color}
\usepackage{longtable}
\PassOptionsToPackage{hyphens}{url}\usepackage{hyperref}
\usepackage{breakurl}
\usepackage{latexsym}
\usepackage{amsmath}
\usepackage{amsthm}
\usepackage{amsfonts}
\usepackage{amssymb}
\usepackage{dsfont}
\usepackage{bm}

\begin{document}

\title{$D_{s0}^*(2317)$ Meson  and $D$-Meson-Kaon Scattering from Lattice QCD }

\author{Daniel Mohler}
\email{dmohler@fnal.gov}
\affiliation{Fermi National Accelerator Laboratory, Batavia, Illinois 60510-5011, USA}

\author{C.~B.~Lang}
\email{christian.lang@uni-graz.at}
\affiliation{Institute of Physics,  University of Graz, A--8010 Graz, Austria}

\author{Luka Leskovec}
\email{luka.leskovec@ijs.si}
\affiliation{Jozef Stefan Institute, 1000 Ljubljana, Slovenia}

\author{Sasa Prelovsek}
\email{sasa.prelovsek@ijs.si}
\affiliation{Department of Physics, University of Ljubljana, 1000 Ljubljana , Slovenia}
\affiliation{Jozef Stefan Institute, 1000 Ljubljana, Slovenia}

\author{R.~M.~Woloshyn}
\email{rwww@triumf.ca}
\affiliation{TRIUMF, 4004 Wesbrook Mall Vancouver, BC V6T 2A3, Canada}

\date{\today}

\begin{abstract}
The scalar  meson $D_{s0}^*(2317)$  is found $37(17)$MeV below $DK$ threshold in a lattice simulation of  the $J^P=0^+$ channel using, for the first time, both $DK$ as well as $\bar sc$ interpolating fields. 
The simulation is done on $N_f=2+1$ gauge configurations with $m_\pi\simeq 156~$MeV, and the resulting $M_{D_{s0}^*}-\tfrac{1}{4}(M_{D_s}+3M_{D_s^*})=266(16)$~MeV is close to the experimental value $241.5(0.8)~$MeV.  The 
energy level related to the scalar meson is accompanied by additional discrete levels due to $DK$ scattering states. The levels near threshold lead to the negative $DK$ scattering length $a_0=-1.33(20)$~fm that indicates the presence of a state below threshold.  
\end{abstract}


\maketitle

The experimentally observed meson spectrum contains a number of states close to an $s$-wave threshold. Such structures are seen in the spectrum of light-quark mesons, in heavy-light mesons and in the spectra of charmonium and bottomonium, and many of them do not fit well with expectations based on a simple quark-antiquark picture. These states are frequently interpreted as either shallow-bound molecules, tightly-bound tetraquarks or mesons containing gluonic excitations and are sometimes referred to collectively as ``exotics''.

In this Letter we focus on the spectrum of charmed-strange $J^P=0^+$ mesons. Prior to the discovery of the $D_{s0}^*(2317)$, quark models predicted a resonance above the D-meson kaon ($DK$) threshold. However, results obtained by various experiments \cite{pdg12} show a very narrow state which is well below the $DK$ threshold. The coupling of $J^P=0^+$ $\bar sc$ to the $DK$ threshold was suggested as a mechanism for lowering the mass of the physical state \cite{vanBeveren:2003kd}. The dependence on $m_\pi$ of the mass differences between the scalar and pseudoscalar heavy-light mesons  was investigated in \cite{Becirevic:2004uv}.
Interestingly, the mass of the $D_{s0}^*(2317)$ is very close to the mass of its non-strange partner, which has given rise to a tetraquark
interpretation \cite{Dmitrasinovic:2005gc}.

Lattice QCD (LQCD) provides the possibility of calculating the spectrum of QCD without resorting to model assumptions. Nevertheless addressing states close to $s$-wave thresholds turns out to be a formidable task. 
In the channel of interest,  $D_{s0}^*(2317)$  needs to be distinguished from the nearby level at $E\simeq m_D+m_K$  (corresponding to the infinite volume $DK$ threshold) which is expected to be shifted on the lattice due to interactions \cite{Luscher:1986pf}. 
Indeed, all physical states with $J^{P}=0^{+}$ and $I=0$ appear as discrete energy levels due to the finite volume.  
The energies of the $s$-wave discrete scattering states $D(\mathbf{p}) K(-\mathbf{p})$, where  $\mathbf{p}\!=\!\tfrac{2\pi}{L}\mathbf{n}$ in the non-interacting limit, are shifted in the the finite volume due the interaction. The energy shifts are related to the infinite volume scattering phase shifts in the elastic region via  L\"uscher's formula \cite{Luscher:1990ux}. 

It was noticed in previous LQCD calculations, that many expected scattering levels are absent \cite{Engel:2010my,Bulava:2010yg,Morningstar:2011ka,Dudek:2010wm} when simulating mesons with quark-antiquark interpolating fields or baryons with three-quark interpolators, although there might be hints of these states in the case of close-by $s$-wave thresholds \cite{Mahbub:2012ri,Engel:2013ig,Alexandrou:2013fsu}. To ameliorate this problem better overlap with multi-particle states is needed which may be obtained by including them explicitly in the basis of lattice interpolators. 

Previous lattice studies \cite{Hein:2000qu,Boyle:1997aq,Boyle:1997rk,Bali:2003jv,Namekawa:2011wt,Moir:2013ub,Mohler:2011ke,Kalinowski:2013wsa,Bali:2012ua,Bali:2011dc} 
considered $D_{s0}^*(2317)$ using only quark-antiquark interpolators. 
Early (quenched) lattice QCD calculations found energy levels substantially above the physical $DK$ threshold \cite{Hein:2000qu,Boyle:1997aq,Boyle:1997rk,Bali:2003jv}. Recent dynamical LQCD simulations  \cite{Namekawa:2011wt,Moir:2013ub,Mohler:2011ke,Kalinowski:2013wsa,Bali:2012ua,Bali:2011dc} are also not definitive due to closeness of the $DK$ threshold.

 \begin{table*}[tbh]
\begin{ruledtabular}
\begin{tabular}{cccccccc}
ID & $N_L^3\times N_T$ & $N_f$ & $a$[fm] & $L$[fm] & \#configs & $m_\pi$[MeV] & $m_K$[MeV]\\ 
\hline
(1)& $16^3\times32$ & 2 & 0.1239(13) & 1.98     & 279 & 266(3)(3) & 552(2)(6)\\
(2)& $32^3\times64$ & 2+1 & 0.0907(13) & 2.90 & 196 & 156(7)(2) & 504(1)(7)\\
\end{tabular}
\end{ruledtabular}
\caption{\label{ensembles}Details of gauge configurations used. $N_L$
and $N_T$ denote the number of lattice points in spatial and time directions, $N_f$  the number of dynamical flavors and $a$ the lattice spacing. The pion mass for ensemble (2) is taken from \cite{Aoki:2008sm}, while the kaon mass results from our calculation with partially quenched strange quarks.}
\end{table*}

A novel feature of the present study is that quark-antiquark and meson-meson interpolators are combined in the charmed-strange $J^P=0^+$ channel. The simulation is performed on two lattice ensembles with parameters listed in   Table \ref{ensembles}. Further details about the ensembles may be found in \cite{Aoki:2008sm} and \cite{Hasenfratz:2008fg,Hasenfratz:2008ce,Lang:2011mn} respectively. To minimize heavy-quark discretization effects at finite lattice spacing the Fermilab method \cite{ElKhadra:1996mp,Oktay:2008ex} is used for the charm quarks. Details of the procedure along with the relevant parameters for ensemble (1) can be found in \cite{Mohler:2012na}. In particular we use the dispersion relations (3) and (18) of \cite{Mohler:2012na} for our tuning procedure and the determination of $p$. The same procedure has been used to tune the charm quark mass on ensemble (2) and the resulting parameters are $c_{sw}=1.64978$ and $\kappa_c=0.12686$. Notice that at finite lattice spacing the mass splittings between states involving a heavy quark are expected to be close to physical while the rest masses are affected by large discretization effects.\footnote{For the rest and kinetic mass of the D meson on ensemble (1) please refer to the caption of Figure (3) in \cite{Mohler:2012na}. For ensemble (2) we obtain $M_1\approx 0.7534$, $M_2\approx 0.828$ and $M_4\approx 0.889$.} Consequently, mass splittings will be quoted with respect to the mass of the spin-averaged 1S state $M_{\overline{1S}}=(M_{D_s}+3M_{D_s^*})/4$ in our final results and in all figures. For ensemble (2) the strange quark mass used in \cite{Aoki:2008sm} differs significantly from the physical value. We therefore use a partially quenched strange quark $m_{s}^{val}\ne m_{s}^{sea}$ and determine the hopping parameter $\kappa_s^{val}$ by minimizing the difference of the $\phi$ meson mass from the experimental mass and the difference of the unphysical $\eta_s$ meson from the value expected from a high-precision lattice determination \cite{Dowdall:2013rya}. The determinations agree to high precision and $\kappa_s=0.13666$ is obtained.

To handle the backtracking quark loops appearing in the Wick contractions, the powerful distillation method \cite{Peardon:2009gh} is used. This can be seen as a smearing prescription producing quark sources and sinks that are approximately Gaussian.  The method allows for a large freedom in the choice of interpolators and for momentum projection at source and sink.
The exact Laplacian-Heaviside version is used for ensemble (1) and the stochastic extension of distillation \cite{Morningstar:2011ka} for ensemble (2). Within this approach we calculate the correlation matrix  
\begin{align}
\label{c}
C_{ij}(t)&=\sum_{t_i}\langle 0|O_i(t_i+t)O_j^\dagger(t_i)|0\rangle\\
&=\sum_n\mathrm{e}^{-tE_n} \langle 0|O_i|n\rangle \langle n|O_j^\dagger|0\rangle\;,\nonumber
\end{align}
using interpolating fields $O_i$ with $J^P=0^+$ (irrep $A_1^+$), isospin $I=0$ and  
 total momentum zero. Four quark-antiquark interpolators $O_{1-4}^{\bar sc}=\bar s A_{1-4} c$ taken to be the entries 1-4 of irrep $A_1^{+}$ in Table XII of \cite{Mohler:2012na} are used. There are also three meson-meson interpolators 
\begin{align}
\label{O}
O_1^{DK}&=\left[\bar{s}\gamma_5u\right](p=0)\left[\bar{u}\gamma_5c\right](p=0)+\left\{u\rightarrow d\right\}\;,\nonumber\\
O_2^{DK}&=\left[\bar{s}\gamma_t\gamma_5u\right](p=0)\left[\bar{u}\gamma_t\gamma_5c\right](p=0)+\left\{u\rightarrow d\right\}\;,\nonumber\\
O_3^{DK}&=\!\!\!\!\!\!\!\!\!\sum_{p=\pm e_{x,y,z}~2\pi/L}\!\!\!\!\!\!\!\left[\bar{s}\gamma_5u\right](p)\left[\bar{u}\gamma_5c\right](-p)+\left\{u\rightarrow d\right\}\;.
\end{align}

\begin{figure}[tb]
\includegraphics[width=85mm,clip]{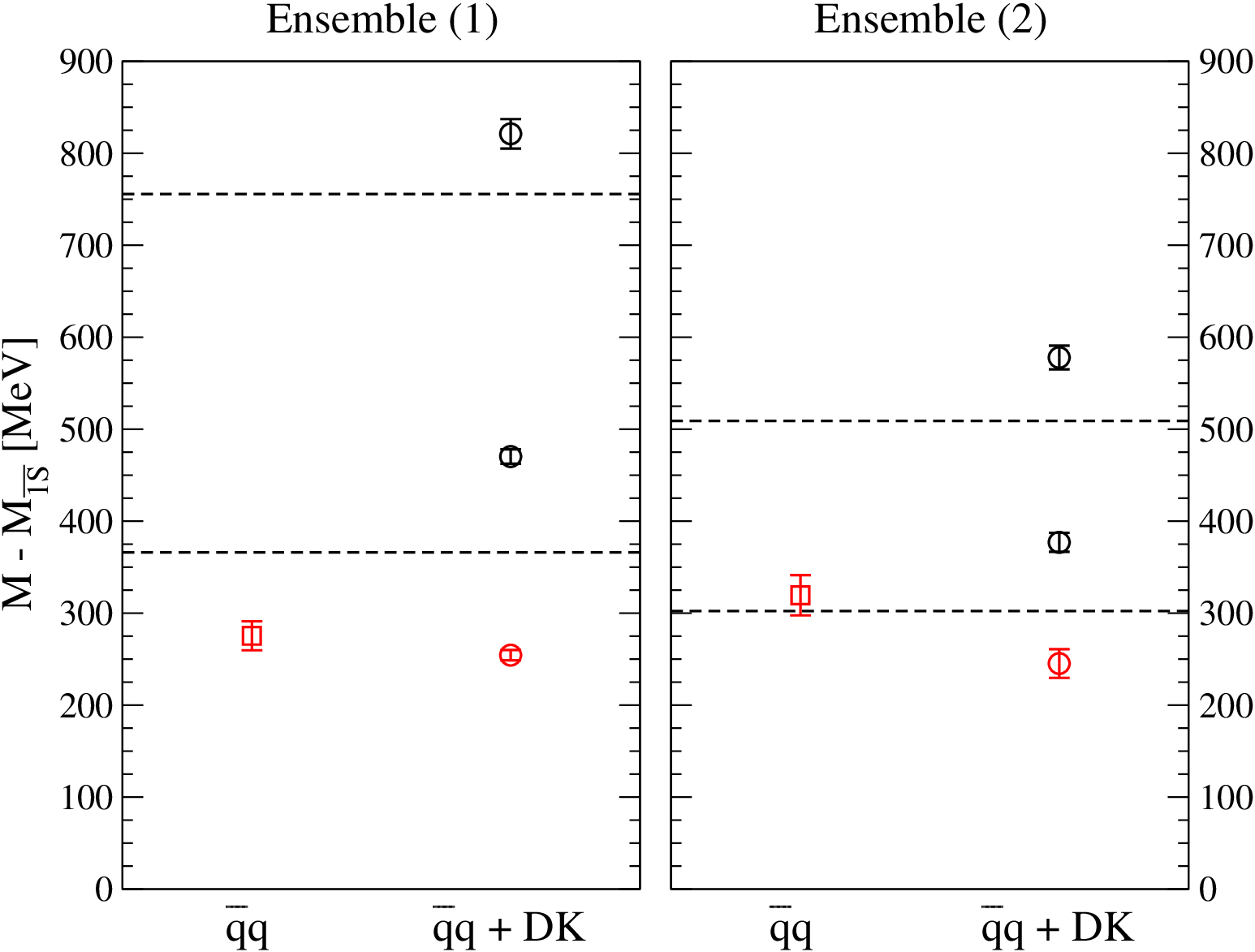}
\caption{ Energy levels from ensemble (1) (left panel) and ensemble (2) (right panel). For each case results with just quark-antiquark ($\bar{q}q$) and with a combined basis of $\bar{q}q$ and $DK$ interpolating fields are shown. The lower dashed lines indicate $M_D+M_K$ on both ensembles, while upper dashed lines show the energies of non-interacting $D(1)K(-1)$. The error bars include statistical and scale setting corrections.} 
\label{fig_levels}
\end{figure}

The discrete energy levels $E_n$ are extracted from the correlators (\ref{c}) using the  variational method \cite{Luscher:1990ck,Michael:1985ne,Blossier:2009kd}.
Figure \ref{fig_levels} illustrates the results for the spectrum obtained from both lattices. 
In each panel the left set of points indicates the ground state level with just a quark-antiquark basis\footnote{The second level from the $\bar qq$ basis is of poor statistical quality and it appears above the second level obtained from the full basis. It is away from the energy region of interest, it does not influence the conclusions and it is not plotted  for clarity.} while the right set of points indicates the energies using our full basis. The lower dashed lines denotes the $m_D+m_K$ threshold on both lattices, while the upper dashed line corresponds to the energy of the non-interacting $D(1)K(-1)$ scattering state.  Note that two low-lying states are observed when using the combined basis. Their signal is unambiguous upon variation of the basis, as long as 
at least one of {\it $O^{DK}_{1,2}$} and at least two $O^{\bar sc}$ interpolators, or if both of {\it $O^{DK}_{1,2}$} and one or more of the 
$O^{\bar sc}$ interpolators 
are used. The interpolator $O^{DK}_3$ is needed to render the $D(1)K(-1)$ state. This level will not be used in the analysis but for our conclusions it is important that it can indeed be identified with the interacting $D(1)K(-1)$. 

Taking a look at the lowest two energy levels on each ensemble there are two possible interpretations:
\begin{enumerate}
\item A sub-threshold state which is stable under the strong interaction (in
  the Isospin limit). We will refer to such a state as a ``bound state'' but
  stress that this choice of words makes no statement about a possible
  $\bar{q}q$ or meson-meson nature of the state. In this case a negative
  scattering length is expected and the up-shifted second level would be
  related to the interacting scattering threshold on the lattice. Such a
  scenario was discussed in the context of a simple model in
  \cite{Sasaki:2006jn}, and was confirmed for a deuteron bound state $pn$
  \cite{Yamazaki:2011nd,Beane:2013br} and for a $D\bar D^*$ bound state
  $X(3872)$ \cite{Prelovsek:2013cra}. The expected behavior of the energy levels in various
scenarios was discussed in model studies in Ref. \cite{MartinezTorres:2011pr}. 
\item A QCD resonance (above threshold). In this case the attraction is not strong enough to form a bound state and the level associated with the finite volume scattering state will be found below threshold. A positive scattering length is expected and the additional level above threshold occurs due to the presence of a resonance in this channel. This is the situation encountered for $D\pi$ scattering in the $J^P=0^+$ channel with resonance $D_0^*(2400)$ \cite{Mohler:2012na} or in $N\pi$ scattering in the negative parity sector \cite{Lang:2012db}.
\end{enumerate}

\begin{figure}[tb]
\includegraphics[width=85mm,clip]{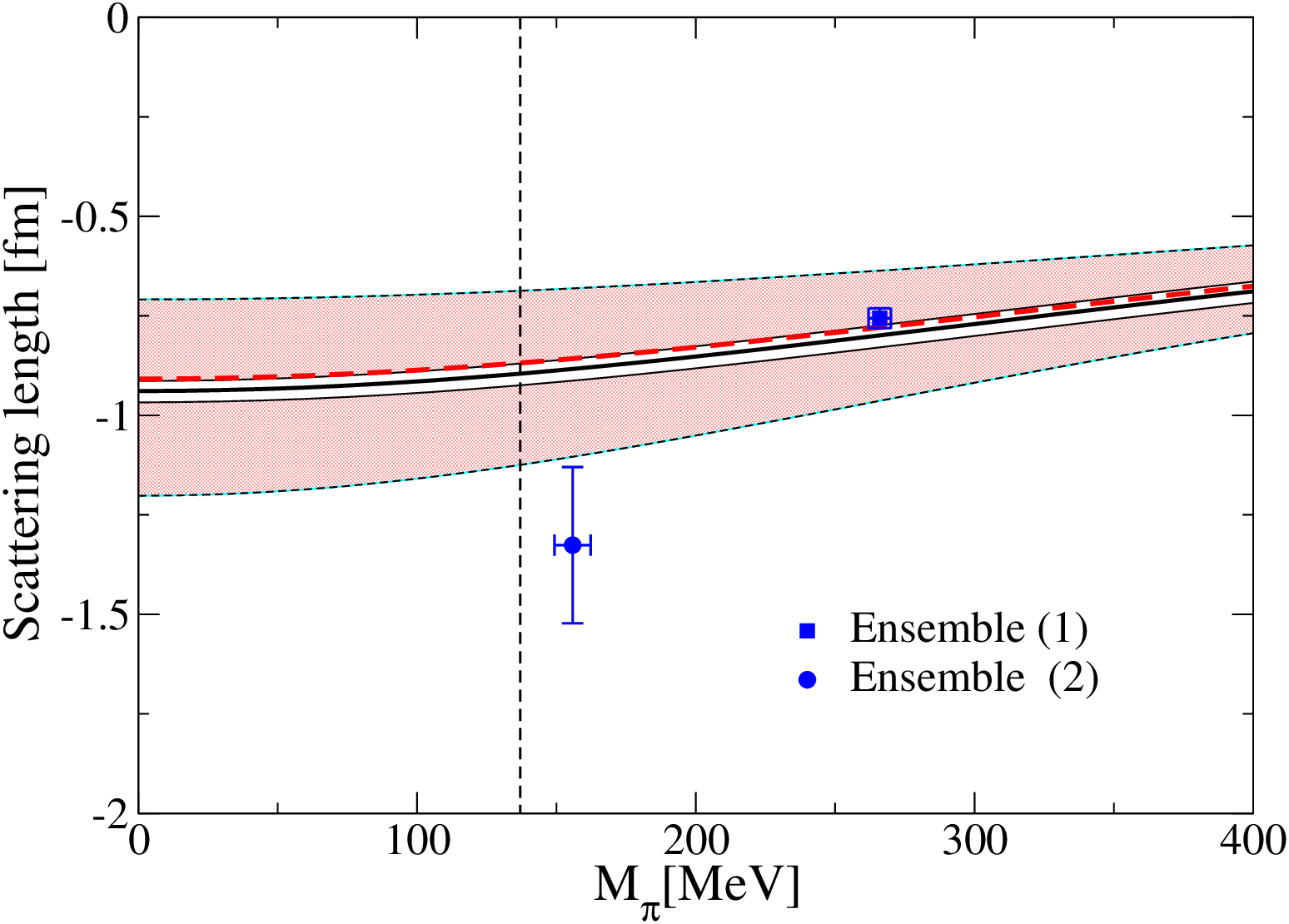}
\caption{ Scattering length $a_0$ for $s$-wave $DK$ scattering with $I=0$. Our result is compared to the expectation from the indirect calculation in \cite{Liu:2012zya}. The vertical line corresponds to the physical pion mass. For an explanation of the curves please refer to the main text.} 
\label{fig_a0}
\end{figure}

The crucial insight is that the plausibility of these scenarios can be tested by determining 
the real number $p\cot\delta$  from energy  levels  using L\"uscher's formula \cite{Luscher:1990ux}, which applies above and below threshold.  Here $\delta$ is the scattering phase shift for the elastic $DK$ scattering in $s$-wave, while $p$ is the $D$ and $K$ momentum related to the energy via $E_{n}(L)=E_D(p)+E_K(p)$. We perform  an effective range approximation  
\begin{equation}
p \cot \delta(p)=\frac{2}{\sqrt{\pi}L}Z_{00}(1,(\tfrac{p\,L}{2\,\pi})^2)\;\approx\frac{1}{a_0}+\frac{1}{2}r_0p^2\;,
\end{equation}
which seems well justified for the momenta at hand (given in Table \ref{values} along with the values for $p\cot\delta$). The lowest two levels give two equations for two unknowns. 

\begin{table}[tb]
\begin{ruledtabular}
\begin{tabular}{cccc}
& level & Ensemble (1) & Ensemble (2)   \\
\hline
$(pa)^2$ & 1 & -0.0347(14)  & -0.0092(24)  \\
   & 2 & 0.0364(14) &  0.0130(16) \\
$(pa)\cot{\delta}$ & 1 & -0.1560(59) & -0.082(19)  \\
   & 2 & -0.1722(74) &  -0.049(15) \\
\hline
$(p_ba)^2$ & - & -0.0250(17) & -0.0060(26)  \\
\end{tabular}
\end{ruledtabular}
\caption{\label{values}Values for $p^2$ and $p\cot{\delta}$ obtained from the two lowest energy levels for both ensembles. In addition the values for the binding momentum $p_b$ are tabulated in the last row.}
\end{table}

For ensemble (1) we obtain\footnote{The effective range value for ensemble (2) has sizeable systematic uncertainty allowing even small negative values but has little influence on the final value of the binding energy.}\vspace{-6pt}
\begin{equation}
a_0=-0.756(25)\;\mathrm{fm} \qquad r_0= -0.056(31)\;\mathrm{fm}\;,
\end{equation}
while ensemble (2) yields
\begin{equation}
a_0=-1.33(20)\;\mathrm{fm} \qquad r_0=  0.27(17)\;\mathrm{fm}\;.
\end{equation}
In both cases the extracted scattering length is negative with a reasonably small statistical and systematic uncertainty, and the effective range is small. This is compatible with scenario (1) above, where the lower energy level is associated with a bound state up to corrections related to the finite volume of the simulation.

Figure \ref{fig_a0} compares our results for the scattering length to the prediction from \cite{Liu:2012zya} where the authors performed a lattice calculation in a variety of channels and extracted the relevant low-energy constants of the chiral effective field theory. These low energy constants were then used to predict the $DK$ $I=0$ scattering length. Two distinct determinations of low energy constants were performed. For the first set of results only the lattice data was used as input (larger error band in the figure, central value thick dashed curve) while for the second set 
the $D_{s0}^*(2317)$ mass from experiment was used as further input leading to much smaller uncertainties (narrower error band in
Fig. \ref{fig_a0}, central value thick solid curve).
As can be seen, our results from the direct simulation agree qualitatively, although the uncertainty close to physical pion and kaon mass is large. 

In infinite volume a bound state would correspond to a pole of the S-matrix which translates to the pole condition $\cot\delta(p_{b})=\mathrm{i}$, where $p_b=\mathrm{i} |p_b|$ denotes the binding momentum of the bound state. Taking the values for $a_0$ and $r_0$ extracted within the effective range approximation, we determine the binding momentum, which translates to our estimate of the bound state energy $M_{L\rightarrow\infty}=E_D(p_b)+E_K(p_b)$ using the dispersion relations $E_K(p)=(M_K^2+p^2)^{1/2}$ and $E_D(p)$ given by Eq. (3) of \cite{Mohler:2012na} with $W_4=0$. It is this bound state energy and its value with respect to the $DK$ threshold that should be compared to experiment. Analogous extraction of the deuteron binding energy from two lowest levels on a single volume was proposed for future simulations in \cite{Beane:2010em}.

The  systematic uncertainties come from fitting the dispersion relation for the $D$-meson \cite{Mohler:2012na} and from determining the kaon mass. For both, the scattering length and the binding energy, we estimate those to be $30\%$ of the statistical errors.

 \begin{table}[tb]
\begin{ruledtabular}
\begin{tabular}{clll}
& $E_1(L)-M_{\overline{1S}}$ & $M^{D_{s0}^*(2317)}_{L\rightarrow\infty}-M_{\overline{1S}}$   \\
\hline
ensemble (1) & 254.4(4.3)(2.3) & 287.2(5.0)(3.0)  \\
ensemble (2) & 245(15)(4) & 266(16)(4) \\
\hline
experiment &   & $241.45(0.60)$  \\
\end{tabular}
\end{ruledtabular}
\caption{\label{masses} The final result for the $D_{s0}^*(2317)$ mass  $M_{L\rightarrow\infty}^{D_{s0}^*(2317)}$, as obtained from the pole condition, compared to the experimental value \cite{pdg12} (right column). The energy levels in the finite-volume lattice are shown in the left column. The errors  are statistical (1st) and due to scale setting (2nd).}
\end{table}

\begin{figure}[t]
\includegraphics[width=86mm,clip]{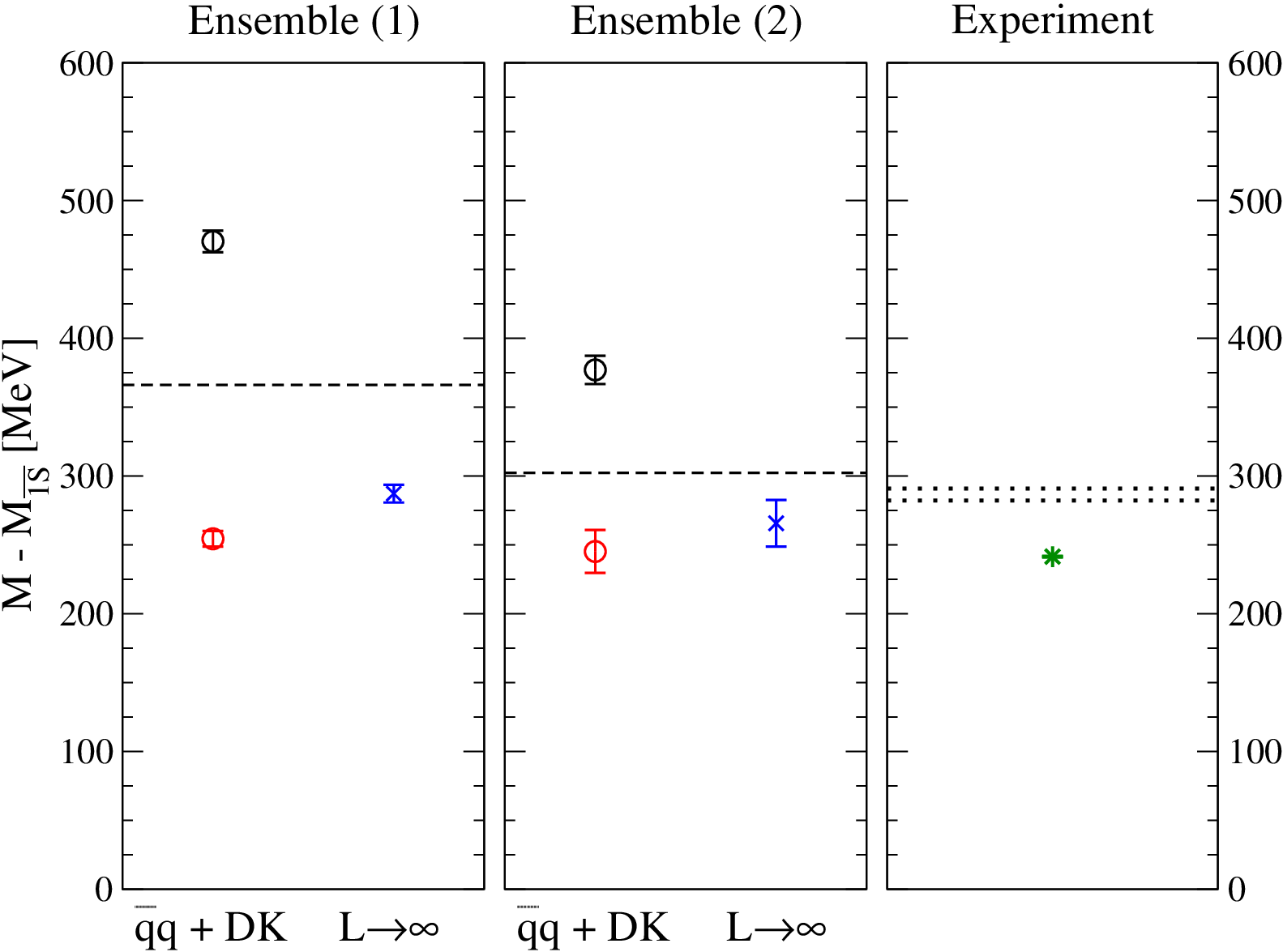}
\caption{The final result for $D_{s0}^*(2317)$ mass is given by the crosses in the left and middle panels, while the experimental value is 
given in the right panel. Instead of the mass itself, we compare the values of  $M_{L\to \infty}^{D_0^*(2317)}-M_{\overline{1S}}$, where 
$M_{\overline{1S}}^{exp}= \tfrac{1}{4}(m_{D_s}+3m_{D_s^*})\simeq 2076$~MeV. The value of the bound state position in the infinite volume limit, $M_{L\to \infty}^{D_0^*(2317)}$ is obtained  from the pole condition $\cot \delta =\mathrm{i}$. The two lowest energy levels from our simulation in the finite volume are given by the circles in the left and middle panels. Dashed lines represent the threshold for $DK$ in our simulation ($m_u=m_d)$, and dotted lines the thresholds for $D^0K^+,~K^0D^+$ in experiment. } 
\label{fig_levels_physical}
\end{figure}
 
Our final result is given alongside the experimental  $D_{s0}^*(2317)$ mass in 
Table \ref{masses} and  Figure \ref{fig_levels_physical}, together with $DK$ thresholds on the lattice and in experiment. Notice that with a pion mass of $156~$MeV and at finite lattice spacing we neither expect the thresholds to agree perfectly, nor do we expect the position of the sub-threshold state to agree exactly with the $D_{s0}^*(2317)$. In particular heavy quark discretization effects of an order of a few percent of the mass splittings are expected and their influence should be addressed in future simulations.

In summary, we have performed a simulation of the $D_s$ $(J^P=0^+)$ spectrum with the novel feature of a combined basis of quark-antiquark and $DK$ operators. The combination of both types of lattice interpolating fields was crucial to obtain energy levels with small statistical uncertainties and the variational analysis shows that both types of operators have
non-vanishing overlap with the physical state. Further notable features of the simulation are the use of an improved heavy-quark action, distillation methods to deal with operator contractions,  and almost physical pions, kaons and D mesons. Unlike previous lattice simulations, we observe a state below $DK$ threshold whose mass is compatible with the experimental $D_{s0}^*(2317)$ within the remaining uncertainties. To obtain precision results, simulations at multiple lattice spacings and with multiple lattice volumes will be needed.

\acknowledgments
We thank Anna Hasenfratz and the PACS-CS collaboration for providing gauge configurations and Martin L\"uscher for making his DD-HMC software available. D.~M. would like to thank E.~Eichten, F.-K. Guo, M.~Hansen, A.~Kronfeld, Y.~Liu and J.~Simone for insightful discussions. The calculations were performed on computing clusters at TRIUMF and the Universities of Graz and Ljubljana. This work is supported in part by the Austrian Science Fund FWF project I1313-N27, by the Slovenian Research Agency ARRS project number N1-0020 and by the Natural Sciences and Engineering Research Council of Canada. Fermilab is operated by Fermi Research Alliance, LLC under Contract No. De-AC02-07CH11359 with the United States Department of Energy. Special thanks to the Institute for Nuclear Theory (University of Washington) for hospitality.


\begin{thebibliography}{10}

\bibitem{pdg12}
Particle Data Group, J.~Beringer {\em et~al.},
\newblock Phys. Rev. {\bf D86}, 010001 (2012).

\bibitem{vanBeveren:2003kd}
E.~van Beveren and G.~Rupp,
\newblock Phys. Rev. Lett. {\bf 91}, 012003 (2003), [arXiv:hep-ph/0305035].

\bibitem{Becirevic:2004uv}
D.~Becirevic, S.~Fajfer and S.~Prelovsek,
\newblock Phys. Lett. {\bf B599}, 55 (2004), [arXiv:hep-ph/0406296].

\bibitem{Dmitrasinovic:2005gc}
V.~Dmitra\ifmmode \check{s}\else \v{s}\fi{}inovi\ifmmode~\acute{c}\else
  \'{c}\fi{},
\newblock Phys. Rev. Lett. {\bf 94}, 162002 (2005).

\bibitem{Luscher:1986pf}
M.~L{\"u}scher,
\newblock Commun. Math. Phys. {\bf 105}, 153 (1986).

\bibitem{Luscher:1990ux}
M.~L{\"u}scher,
\newblock Nucl. Phys. {\bf B354}, 531 (1991).

\bibitem{Engel:2010my}
BGR [Bern-Graz-Regensburg], G.~P. Engel, C.~B. Lang, M.~Limmer, D.~Mohler and
  A.~Sch{\"a}fer,
\newblock Phys. Rev. {\bf D82}, 034505 (2010), [arXiv:1005.1748].

\bibitem{Bulava:2010yg}
J.~Bulava {\em et~al.},
\newblock Phys. Rev. {\bf D82}, 014507 (2010), [arXiv:1004.5072].

\bibitem{Morningstar:2011ka}
C.~Morningstar {\em et~al.},
\newblock Phys. Rev. {\bf D83}, 114505 (2011), [arXiv:1104.3870].

\bibitem{Dudek:2010wm}
J.~J. Dudek, R.~G. Edwards, M.~J. Peardon, D.~G. Richards and C.~E. Thomas,
\newblock Phys. Rev. {\bf D82}, 034508 (2010), [arXiv:1004.4930].

\bibitem{Mahbub:2012ri}
M.~S. Mahbub, W.~Kamleh, D.~B. Leinweber, P.~J. Moran and A.~G. Williams,
\newblock Phys. Rev. {\bf D87}, 011501 (2013), [arXiv:1209.0240].

\bibitem{Engel:2013ig}
G.~P. Engel, C.~B. Lang, D.~Mohler and A.~Schaefer,
\newblock Phys. Rev. {\bf D87}, 074504 (2013), [arXiv:1301.4318].

\bibitem{Alexandrou:2013fsu}
C.~Alexandrou, T.~Korzec, G.~Koutsou and T.~Leontiou,
\newblock {Nucleon Excited States in $N_f$=2 lattice QCD},
\newblock 2013, [arXiv:1302.4410].

\bibitem{Hein:2000qu}
J.~Hein {\em et~al.},
\newblock Phys. Rev. {\bf D62}, 074503 (2000), [arXiv:hep-ph/0003130].

\bibitem{Boyle:1997aq}
UKQCD, P.~Boyle,
\newblock Nucl. Phys. Proc. Suppl. {\bf 53}, 398 (1997).

\bibitem{Boyle:1997rk}
UKQCD, P.~Boyle,
\newblock Nucl. Phys. Proc. Suppl. {\bf 63}, 314 (1998),
  [arXiv:hep-lat/9710036].

\bibitem{Bali:2003jv}
G.~S. Bali,
\newblock Phys. Rev. {\bf D68}, 071501 (2003), [arXiv:hep-ph/0305209].

\bibitem{Namekawa:2011wt}
PACS-CS Collaboration, Y.~Namekawa {\em et~al.},
\newblock Phys. Rev. {\bf D84}, 074505 (2011), [arXiv:1104.4600].

\bibitem{Moir:2013ub}
G.~Moir, M.~Peardon, S.~M. Ryan, C.~E. Thomas and L.~Liu,
\newblock JHEP {\bf 1305}, 021 (2013), [arXiv:1301.7670].

\bibitem{Mohler:2011ke}
D.~Mohler and R.~M. Woloshyn,
\newblock Phys. Rev. {\bf D84}, 054505 (2011), [arXiv:1103.5506].

\bibitem{Kalinowski:2013wsa}
M.~Kalinowski and M.~Wagner,
Acta Phys, Pol. B Proc. Suppl. {\bf 6}, 991 (2013), 
[arXiv:1304.7974].

\bibitem{Bali:2012ua}
G.~Bali, S.~Collins and P.~Perez-Rubio,
\newblock J.Phys.Conf.Ser. {\bf 426}, 012017 (2013), [arXiv:1212.0565].

\bibitem{Bali:2011dc}
G.~Bali {\em et~al.},
\newblock PoS {\bf LATTICE2011}, 135 (2011), [arXiv:1108.6147].

\bibitem{Aoki:2008sm}
PACS-CS, S.~Aoki {\em et~al.},
\newblock Phys. Rev. {\bf D79}, 034503 (2009), [arXiv:0807.1661].

\bibitem{Hasenfratz:2008fg}
A.~Hasenfratz, R.~Hoffmann and S.~Schaefer,
\newblock Phys. Rev. D {\bf 78}, 014515 (2008), [arXiv:0805.2369].

\bibitem{Hasenfratz:2008ce}
A.~Hasenfratz, R.~Hoffmann and S.~Schaefer,
\newblock Phys. Rev. D {\bf 78}, 054511 (2008), [arXiv:0806.4586].

\bibitem{Lang:2011mn}
C.~B. Lang, D.~Mohler, S.~Prelovsek and M.~Vidmar,
\newblock Phys. Rev. {\bf D84}, 054503 (2011), [arXiv:1105.5636].

\bibitem{ElKhadra:1996mp}
A.~X. El-Khadra, A.~S. Kronfeld and P.~B. Mackenzie,
\newblock Phys. Rev. {\bf D55}, 3933 (1997), [arXiv:hep-lat/9604004].

\bibitem{Oktay:2008ex}
M.~B. Oktay and A.~S. Kronfeld,
\newblock Phys. Rev. {\bf D78}, 014504 (2008), [arXiv:0803.0523].

\bibitem{Mohler:2012na}
D.~Mohler, S.~Prelovsek and R.~M. Woloshyn,
\newblock Phys. Rev. {\bf D87}, 034501 (2013), [arXiv:1208.4059].

\bibitem{Dowdall:2013rya}
R.~J. Dowdall, C.~T.~H. Davies, G.~P. Lepage and C.~McNeile,
\newblock {$V_{us}$ from $\pi$ and $K$ decay constants in full lattice QCD with
  physical $u$, $d$, $s$ and $c$ quarks},
\newblock 2013, [arXiv:1303.1670].

\bibitem{Peardon:2009gh}
Hadron Spectrum Collaboration, M.~Peardon {\em et~al.},
\newblock Phys. Rev. D {\bf 80}, 054506 (2009), [arXiv:0905.2160].

\bibitem{Luscher:1990ck}
M.~L{\"u}scher and U.~Wolff,
\newblock Nucl. Phys. {\bf B339}, 222 (1990).

\bibitem{Michael:1985ne}
C.~Michael,
\newblock Nucl. Phys. {\bf B259}, 58 (1985).

\bibitem{Blossier:2009kd}
B.~Blossier, M.~Della~Morte, G.~von Hippel, T.~Mendes and R.~Sommer,
\newblock JHEP {\bf 04}, 094 (2009), [arXiv:0902.1265].

\bibitem{Sasaki:2006jn}
S.~Sasaki and T.~Yamazaki,
\newblock Phys. Rev. {\bf D74}, 114507 (2006), [arXiv:hep-lat/0610081].

\bibitem{Yamazaki:2011nd}
Collaboration for the PACS-CS, T.~Yamazaki, Y.~Kuramashi and A.~Ukawa,
\newblock Phys. Rev. {\bf D84}, 054506 (2011), [arXiv:1105.1418].

\bibitem{Beane:2013br}
S.~Beane {\em et~al.},
Phys. Rev. C {\bf 88}, 024003 (2013),
\newblock 2013, [arXiv:1301.5790].

\bibitem{Prelovsek:2013cra}
S.~Prelovsek and L.~Leskovec,
\newblock Phys. Rev. Lett. {\bf 111}, 192001 (2013), [arXiv:1307.5172].

\bibitem{MartinezTorres:2011pr}
A.~Mart{\'\i}nez~Torres, L.~R. Dai, C.~Koren, D.~Jido and E.~Oset,
\newblock Phys. Rev. D {\bf 85}, 014027 (2012), [arXiv:1109.0396].

\bibitem{Lang:2012db}
C.~B. Lang and V.~Verduci,
\newblock Phys. Rev. D{\bf 87}, 054502 (2013), [arXiv:1212.5055].

\bibitem{Liu:2012zya}
L.~Liu, K.~Orginos, F.-K. Guo, C.~Hanhart and U.-G. Meissner,
\newblock Phys. Rev. D {\bf 87}, 014508 (2013), [arXiv:1208.4535].

\bibitem{Beane:2010em}
 S. R. Beane, W. Detmold, K. Orginos, and M. J. Savage,
Prog. Part. Nucl. Phys. {\bf 66}, 1 (2011), [arXiv:1004.2935].

\end{thebibliography}

\end{document}